# First-principles investigation of elastic anomalies in niobium at high pressure and temperature

Yi X. Wang,[1,2] Hua Y. Geng,[2,*] Q. Wu,[2] Xiang R. Chen,[1,*] and Y. Sun[2]

[1]Institute of Atomic and Molecular Physics, College of Physical Science and Technology, Sichuan University, Chengdu 610064, China;

[2] National Key Laboratory of Shock Wave and Detonation Physics, Institute of Fluid Physics, CAEP, Mianyang 621900, China.

**Abstract:** Niobium does not show any structure transition up to very high pressures. Nonetheless, by using density functional theory, we demonstrate in this work that it exhibits striking softening in elastic moduli $C_{44}$ and $C'$ at a pressure from 20 to 150 GPa. A novel anomaly softening in $C_{44}$ from 275 to 400 GPa is also predicted. The physics behind these two anomalies is elaborated by electronic structure calculations, which revealed that they are actually different, with the first one directly relates to an underlying rhombohedral distortion whereas the latter originates in an electronic topological transition. The large magnitude of the softening leads to a remarkable elastic anisotropy in both the shear and the Young's moduli of Nb. Further investigation shows that thermo-electrons have an important role on these anomalies. This effect has not been noticed before. With increased electronic temperature, it is found that all anomalies (both the elastic softening and anisotropy) in Nb are gradually diminished, effectively giving rise to a temperature-induce hardening phenomenon.

**Keywords:** transition metal, elastic anomalies, first-principles, high pressure and temperature

# I. INTRODUCTION

---

*Correspondence and requests for materials should be addressed to H.-Y. G. (e-mail: s102genghy@caep.ac.cn) or X.-R. C. (e-mail: xrchen@scu.edu.cn).

As a 4*d* transition metal with a body-centered cubic (BCC) structure, niobium has attracted considerable attention[1,2] due to its importance in electronic, nuclear, and superconducting applications. For example, Nb has the highest superconducting transition temperature of $T_c$ = 9.25 K among the elemental metals at ambient pressure.[3] It is also one of the most promising refractory metal due to its high melting point and good ductility at room temperature.[4,5] The equations of state (EOS) of Nb have been studied by some experimental works,[6-10] and none of them show any sign of discontinuity up to ~150 GPa. This indicates that Nb does not have a structural transformation with large volume change in this pressure range. Theoretical calculations also suggested that the BCC phase of Nb is stable up to very high pressures.[11] Despite its simple crystalline structure, Nb demonstrates unexpected anomalies in both mechanical and electronic properties. Singh *et al.*[12] studied the compressive strength of Nb under high pressures, and found that the strength exhibits an anomalous variation between 0 and 15 GPa, which then increases almost linearly with the pressure and reaches a value of ~0.94 GPa at 40 GPa. The effect of pressure on the superconducting transition temperature of Nb has also been studied,[3] which revealed two subtle anomalies at 5 and 60 GPa, respectively: $T_c$ jumps from 9 to 10 K at 5 GPa, it remains nearly as a constant up to 60 GPa, and then falls off at higher pressures. This anomaly has been attributed to electronic topological transition (ETT),[13] which involves a topological change in the Fermi surface driven by compression.

Furthermore, the anomaly in the elastic constants of Nb also attracted a lot of attentions. To our best knowledge, the first indication of the elastic anomaly in Nb perhaps came from the experimental observation by Nakagawa *et al.*[14] They found an anomalous softening in phonon dispersions of Nb along [00ζ] direction at ζ = 0.2. Recently, Landa *et al.*[15] confirmed that discovery with first-principles calculations by showing that the shear elastic constants $C_{44}$ and $C'$ of Nb soften at a pressure of ~50 GPa. They suggested that this pressure-induced shear softening originates from the electronic structure with Fermi surface (FS) nesting. Moreover, Koči *et al.*[16] studied the elasticity of V, Nb, Ta, Mo, and W at high pressures by using first-principles calculations. They found that the $C_{11}$, $C_{12}$, and $C_{44}$ moduli for the group VB elements all display an erratic behavior, and the Fermi surface analysis revealed a shrinking nesting vector with pressure, with the nesting vector reduces to zero at 247, 74, and 275 GPa for V, Nb, and Ta, respectively.

Recently, a static compression experiment based on synchrotron x-ray diffraction by Ding *et al*.[17] unexpectedly discovered a transition in V from the BCC structure to a rhombohedral (RH) phase at ~69 GPa. This finding is remarkable, since previous understanding suggested that the BCC phase of vanadium should be the most stable structure up to at least 154 GPa.[8] A very recent theoretical work further revealed that this transition is sensitively affected by the localization of the *d* electrons, and can be easily modified by thermo-electrons.[18] This discovery inspires the speculation that the same mechanism might also present in Nb, and could probably take the responsibility for the anomalous properties of Nb, since they have almost the same electronic structure.

As mentioned out above, the underlying mechanism for the anomalous softening in Nb is still under debate. For example, the ETT in Nb occurs at ~110 GPa, the band Jahn-Teller effect becomes effective at about 100 GPa, and the FS nesting covers a narrow pressure range of 0~75 GPa. However, none of them along is fully consistent with the $C_{44}$ softening that spans from 20 to 150 GPa. How does this softening relate to what occurs in V and Ta is still unclear. Furthermore, most investigations on this problem mainly focused on the low pressure range.[15,19,20] There was very few study devoted to higher pressures.[16] The temperature effect on the mechanical properties is also completely ignored, which, however, might be very important.[18] In order to acquire a thorough and profound comprehension about this topic, we revisit the elastic properties of Nb at high pressure and high temperature in this work. Intriguing results have been found. A novel softening at high pressure is predicted for the first time. The underlying physical mechanisms of these anomalies have been elucidated. The paper is organized as follows. The details of the computational method are described in Sec. II. The results and discussion are presented in Sec. III and IV, respectively. Finally, the summary of the main conclusions are given in Sec. V.

## II. THEORETICAL METHOD AND COMPUTATIONAL DETAILS

The theoretical simulations are performed using the Vienna *Ab initio* Simulation Package (VASP),[21] which is based on the density functional theory (DFT)[22] and the projector augmented-wave (PAW) method.[23] The PAW pseudopotential contains 13 valence electrons (including $4s^2$, $4p^6$, $4d^4$, and $5s^1$ states). The Perdew-Burke- Ernzerhof (PBE) generalized gradient approximation (GGA)[24] for the electronic exchange-correlation functional is used. The kinetic energy cut-off of the plane waves is taken as 900 eV, and a $60 \times 60 \times 60$ uniform mesh is used for the *k*-point sampling. This parameter setting is carefully checked by increasing the cutoff energy and *k*-points to higher values to ensure that it gives an absolutely converged total energy and pressure.

To evaluate the elastic constants of the BCC phase, we use a conventional 2-atom cubic unit cell to calculate the total energy as a function of volume and the distortion along shear strains. The shear moduli $C_{44}$ and $C' = (C_{11}-C_{12})/2$ are then obtained from the second derivatives of the total energy with respect to the deformation magnitude $\delta$, which is defined in the strain matrices as[19]

$$\varepsilon_{C_{44}} = \begin{pmatrix} 0 & \delta & 0 \\ \delta & 0 & 0 \\ 0 & 0 & \delta^2/\left(1-\delta^2\right) \end{pmatrix}, \tag{1}$$

$$\varepsilon_{C_{11}-C_{12}} = \begin{pmatrix} \delta & 0 & 0 \\ 0 & -\delta & 0 \\ 0 & 0 & \delta^2/\left(1-\delta^2\right) \end{pmatrix}. \tag{2}$$

The corresponding strained energies are then expressed as

$$E\left(\delta\right) = E\left(0\right) + 2C_{44}V\delta^2 + O\left(\delta^4\right), \tag{3}$$

$$E\left(\delta\right) = E\left(0\right) + \left(C_{11} - C_{12}\right)V\delta^2 + O\left(\delta^4\right). \tag{4}$$

The other two elastic constants $C_{11}$ and $C_{12}$ are then derived by using the calculated bulk modulus

$$B = -V\left(\frac{\partial P}{\partial V}\right)_T = \frac{1}{3}\left(C_{11} + 2C_{12}\right). \tag{5}$$

As will be shown below, there is a large difference in the numeric value between $C_{44}$ and $C'$. This indicates that the elastic response of a single crystal Nb is highly anisotropic. For cubic crystals, Zener[25] introduced a measure for this anisotropy:

$$A = \frac{2C_{44}}{C_{11} - C_{12}}. \tag{6}$$

For an isotropic crystal, this shear anisotropy $A$ always equals to 1. Other values suggest the anisotropy degree of the crystal.[26]

Zener anisotropy only applies to cubic crystals. In order to measure the anisotropy for all crystals with a single-valued quantity, Chung and Buessem[27] proposed the following empirical measure

$$A_G = \frac{G_V - G_R}{G_V + G_R} = \frac{3(A-1)^2}{3(A-1)^2 + 25A}, \tag{7}$$

where $G_V$ and $G_R$ are the shear modulus' Voigt and the Reuss estimates, respectively. With this expression, a value of zero represents the complete elastic isotropy, and a value of 1 indicates the largest anisotropy.

Ranganathan and Ostoja-Starzewshi[28] alternatively took all stiffness coefficients into account rather than only their ratios like Zener's or Chung and Buessem's definition by recognizing the tensorial nature of the elastic stiffness to define the anisotropy. This universal anisotropy index for a single crystal can be written as

$$A_U = \frac{5G_V}{G_R} + \frac{B_V}{B_R} - 6. \tag{8}$$

For locally isotropic single crystals, $A_U$ is identical to zero. The departure of $A_U$ from zero characterizes the magnitude of anisotropy. For a cubic crystal system, the $G_V$, $G_R$, $B_V$, and $B_R$ are given by

$$G_V = \frac{C_{11} - C_{12} + 3C_{44}}{5}, \tag{9}$$

$$G_R = \frac{5(C_{11} - C_{12})C_{44}}{4C_{44} + 3(C_{11} - C_{12})}, \tag{10}$$

$$B_V = B_R = \frac{C_{11} + 2C_{12}}{3}. \tag{11}$$

Besides these standard shear deformations, we also study the RH distortions, inspired by the BCC→RH structural transition observed in vanadium. The volume-conserved BCC→RH distortion matrix is defined as

$$T(\delta) = \begin{pmatrix} k & \delta & \delta \\ \delta & k & \delta \\ \delta & \delta & k \end{pmatrix}, \qquad (12)$$

in which $k$ is determined from the real positive solution of $\det(T) = 1$, to ensure a volume-conserving transformation. The small displacement $\delta$ represents the amount of the rhombohedral deformation in the BCC crystalline system. The details of this distortion are referenced to Refs. 29 and 18.

## III. RESULTS

### A. Structural properties and equation of state

Before discussing the anomalous mechanical properties, it is necessary to assess the structure and EOS of Nb at high pressures. We adopt the following procedure. First, we take a series of different values of lattice constant $\varepsilon$ to calculate the total energy $E$ and the corresponding primitive cell volume $V$, and then optimize the crystal structure (with a supercell) to obtain the lowest energy $E_{min}$ for the given $\varepsilon$. This procedure is repeated over a wide range of $\varepsilon$. Finally, by fitting the data $E_{min}$-$V$ to the Vinet equation of state[30]

$$\ln\left(\frac{Px^2}{3(1-x)}\right) = \ln B_0 + a(1-x), \qquad (13)$$

in which

$$x = \left(\frac{V}{V_0}\right)^{1/3}, \quad a = \frac{3}{2}\left(B_0' - 1\right), \qquad (14)$$

we acquire the equilibrium primitive cell volume $V_0$, the bulk modulus $B_0$, and its pressure derivative $B_0'$ at 0 K and 0 GPa. The obtained results are listed in Table I, together with available experimental[9,31] and theoretical[16] data. It can be seen that our results are in good agreement with the experimental data, and a little bit better than previous theoretical estimate as reported in Ref. 16. This might be due to the coarse $k$-point grid of 30×30×30 they used. Especially, our calculated equilibrium volume $V_0$ and bulk modulus $B_0$ are within 0.72% and 0.83% of the experimental values, respectively. In addition, we also calculated the EOS of Nb at zero Kelvin up to 400

GPa, as shown in Fig. 1. It is evident that our calculated pressure-volume curve does not show any anomaly. It is fully consistent with the reduced shock wave experimental results[9,32] and previous theoretical data.[15] The slight deviation in the DAC data could stem from the non-hydrostaticity in the experiment or errors in the pressure scale they used.[9] This assessment validates our method, and demonstrates that our parameter setting is accurate enough for the following mechanical property study.

## B. Anomalies in elastic moduli at zero Kelvin

The EOS discussed above is a spherically averaged quantity. The underlying anisotropy has been smoothed out. In order to gain a full comprehension of high pressure behavior of this transition metal, we further investigate the elastic properties of Nb. The elasticity of a material is totally characterized by its elastic stiffness tensor, which are the coefficients that relate the stress tensor to the strain tensor by Hooke's law. Since the stress and strain tensors are symmetric, the most general elastic stiffness tensor has only 21 independent components. For a cubic crystal, they are further reduced to only three, i.e., $C_{11}$, $C_{12}$ and $C_{44}$. Our calculated elastic constants of Nb in BCC phase at the ground state, along with the available experimental[33] and other theoretical data,[15,16,19] are given in Table II. It is obvious that our $C_{11}$ and $C_{12}$ are in good agreement with the experimental and theoretical values. For the elastic constant $C_{44}$, our result is better by comparison to the theoretical results of Refs. 16 and 19, but is still underestimated by about 30% if compared to the experimental data. In order to understand the softening in $C_{44}$, Liu *et al.*[19] calculated the density of state (DOS) and the Fermi surface (FS) of Nb using DFT. They argued that the most probable reason for $C_{44}$ underestimation is related to the nesting feature in the FS which produces a Van Hove singularity in the electronic DOS close to the Fermi level. In addition, Koči *et al.*[16] studied the FS by first-principles calculations. They also found that the length of the nesting vector **q** for Nb decreases with pressure and disappears at 74 GPa. The deficiency of DFT when describing the nesting vectors could therefore be a possible reason for the $C_{44}$ underestimation.

The calculated elastic moduli $C_{11}$, $C_{12}$, $C_{44}$, and $C'$ of Nb in BCC phase as a function of pressure at zero Kelvin, along with the theoretical results of Koči et al.[16] and Landa et al.,[15] are plotted in Fig. 2. It is evident that our results are consistent well with these theoretical results within the studied pressure range. This again justifies the adequacy of our method. Moreover, we find that the monoclinic shear modulus $C_{44}$ softens at a pressure range of 20 to 150 GPa. The tetragonal shear modulus $C'$ is also anomalously softening within the same pressure range. For comparison, we also display the measured yield strength of Nb as a function of pressure[12] in the inset of Fig. 2 (b). Similar to the shear elastic moduli $C_{44}$ and $C'$, the measured yield strength is also soften within a pressure range of 0~45 GPa.[12] This strength softening is unusual, because the strength and ductility of metals are usually enhanced by increased hydrostatic compression.[34]

According to previous studies,[35-37] we know that the pressure-induced structural transformation is usually one of the most common reasons for strength loss. However, there is no structural transition in Nb, and the BCC is always the most stable phase. Softening of $C_{44}$ and $C'$ is also observed in vanadium but with a much larger magnitude,[18] where the $C_{44}$ drops to zero and a BCC→RH transition is induced. In Nb, however, the $C_{44}$ is always positive, indicating that the BCC structure is mechanically stable and there is no shear induced structural transition. Based on this, one might argue that the softening of $C_{44}$ is purely due to FS nesting, as previously suggested, and has nothing to do with the rhombohedral distortion as observed in vanadium. Furthermore, it is interesting to notice that though tantalum has a similar electronic structure and thus the same FS nesting, it does not show any softening at the low pressure range. This observation contrasts with the thumb rule that the high pressure phenomenon of a light element will occur at a lower pressure for the heavier elements in the same group. The $C_{44}$ of Ta softens only at higher pressures from 140 to 250 GPa,[16] implying that FS nesting might not be the only driven force for this softening. The anomalous softening of Ta at around 200 GPa also suggests that Nb should have another (yet unknown) softening at higher pressure.

Indeed, our extensive calculation reveals that the $C_{44}$ of Nb also slightly softens at pressures from 275 to 400 GPa, as shown in Fig. 2 (b). This softening has never been reported before. As mentioned above, the FS nesting in Nb has finished at 75 GPa.[20] Thus the softening at this high pressure range should be driven by a mechanism other than the FS nesting.

## C. Anomalous behavior of elastic anisotropy at zero Kelvin

Anomalous softening in both $C_{44}$ and $C'$ implies that Nb should have an extraordinary anisotropy, which has been completely obscured in the EOS. An adequate description of such anisotropic behavior has an important implication in understanding the mechanics of this group metals. The elastic anisotropy is also important in material science, such as phase transformations, dislocation dynamics, and other geophysical applications.[38] It is thus beneficial to quantify the anisotropy of elastic properties of Nb.

In Table III, the calculated shear anisotropy indices $A$, $A_G$, and $A_U$ of Nb in BCC phase under a pressure up to 100 GPa at zero Kelvin are listed. At zero pressure, our calculated $A$, $A_G$, and $A_U$ of Nb are 0.34 (which is consistent with other theoretical assessment of 0.3 in Ref. 19 and 0.32 in Ref. 39), 0.13, and 1.31, respectively. As mentioned above, our calculated $C_{44}$ at 0 GPa is underestimated by about 30% if compared to the experimental data. This gives rise to a lower shear anisotropy index compared to the value of $A = 0.5$ obtained from the measured single-crystal elastic constants.[33] The calculated elastic anisotropy of Nb are therefore slightly overestimated. However, its relative change with increasing pressure should still be reliable.

In order to give an intuitive impression about the anisotropy magnitude of Nb, we compare its anisotropy with other typical cubic metals at ambient conditions. The anisotropy character of Nb is similar to vanadium in the same group, which is 0.39 as reported in Ref. 39 and 0.36 in Ref. 40. Other transition metals are more isotropic, with $A$ of 0.61 for Cr,[40] and 0.66~0.67 for Mo.[40,41] The anisotropy of commonly used metals is quite different, with $A = 2.14$ for Fe,[40] 2.73 for Au,[42] and 3.27 for Cu,[43] for all of them $C_{44} > C'$. The heavier neighbor Ta also has $C_{44} > C'$, and the anisotropy factor $A$ is about 1.37, as reported in Ref. 39. It should be noted that the anisotropy factor $A$ of Nb first decreases with pressure, which reaches a minimal value of 0.16 at 60 GPa, and then increases beyond that pressure. To our best knowledge, $\delta$-Pu shows the largest anisotropy in all face-centered cubic metals, whose shear moduli $C_{44}$ and $C'$ differ by a factor of ~7 in pure Pu measured by ultrasonic[44] and x-ray.[45] Further calculation suggested that addition of Ga into Pu softens the metal further, and the highest anisotropy can be ~9 for 6 at. % Ga, as reported in Ref. 46. In addition, alkali metals also show a high anisotropy, with the highest elastic anisotropy factor up to 8.5 in Li.[41,47] It is interesting to note that our calculated anisotropy magnitude of Nb ($A^{-1} \approx 6$) at 60 GPa is comparable to that of $\delta$-Pu and Li. The high anisotropy implies that Nb lies at the vicinity of a monoclinic shear $C_{44}$-induced Born mechanical instability boundary (where both $\delta$-Pu and Li are close to a tetragonal shear induced instability). Due to the difference in the definition, the variation of $A_G$ and $A_U$ are opposite and they reach a maximum at 60 GPa, where Nb shows the highest anisotropy.

The elastic anisotropy of a material also manifests in the direction-dependent linear compressibility and Young's modulus. The linear compressibility refers to the relative decrease in the length along a given direction when the crystal is subjected to a hydrostatic pressure. The directional dependent Young's modulus $E$ and linear compressibility $\beta$ can be expressed as[48]

$$E = \left( l_1^4 S_{11} + 2l_1^2 l_2^2 S_{12} + 2l_1^2 l_3^2 S_{13} + l_2^4 S_{22} + 2l_2^2 l_3^2 S_{23} + l_3^4 S_{33} + l_2^2 l_3^2 S_{44} + l_1^2 l_3^2 S_{55} + l_1^2 l_2^2 S_{66} \right)^{-1},$$

(15)

$$\beta = \left( S_{11} + S_{12} + S_{13} \right) l_1^2 + \left( S_{12} + S_{22} + S_{23} \right) l_2^2 + \left( S_{13} + S_{23} + S_{33} \right) l_3^2,$$

(16)

where $l_1$, $l_2$ and $l_3$ are the direction cosines, $S_{ij}$ are the elements in the compliance tensor, which is the inverse of the elastic tensor. With these expressions and the elastic constants, we obtain the direction-dependent linear compressibility and Young's modulus for BCC Nb at a pressure of 60 GPa. It shows that the linear compressibility is strictly isotropic, whereas the Young's modulus displays a strong anisotropy, which is in agreement with the well-known behavior of other cubic crystals system.[49,50] To intuitively understand the elastic anisotropy, the anisotropic characters of shear modulus, Young's modulus and linear compressibility of BCC Nb at 60 GPa are also calculated (see Figs. S8 and S9 in Supplementary Material).

## D. Anomalies contributed from thermo-electrons

Similar to its light neighbor V, Nb also has partial localized $d$ electrons just below the Fermi level. Previous work revealed that due to electronic correlations, the electronic entropy contribution from these orbitals is very important even at relatively low temperatures.[18] Therefore thermo-electrons could have a significant impact on the shear moduli and mechanical anisotropy. To explore this interesting postulation, a thorough investigation on the thermo-electron effect has been performed by using the finite temperature DFT calculation, which includes the thermo-electronic entropy explicitly. In Fig. 3, we plot the electronic temperature effects on the shear modulus $C_{44}$ of BCC phase of Nb as a function of pressure at different temperatures. As expected, thermo-electrons increase $C_{44}$ greatly when between 0 and 150 GPa, and the softening is reduced. The enhanced shear modulus indicates the re-stabilization of the BCC phase according to the Born mechanical instability criterion. We notice that this thermo-electron effect only presents in the pressure ranges of 0~175 GPa and 275~400 GPa. At pressure in 175~275 GPa, thermo-electronic effects are negligible. This may be a result of interplay between thermo-electronic entropy and underlying instability of the BCC phase of Nb. Therefore, though Nb is different from V and does not have a BCC→RH transition, it also has the exotic temperature-induced *hardening* as predicted for V.[18] It is the first time that such a bizarre phenomenon is predicted for Nb. Explicitly, at 0 GPa and 75 GPa, the $C_{44}$ of Nb increases by 12.9% and 135.3% when the temperature is increased from 0 to 2000 K, respectively. The melting temperature of Nb at ambient pressure is 2750.6 K.[51] Compression to 75 GPa would increase it to about 4600 K, if estimated using the slope of the melting curve of Ta.[52] The variation of $C_{44}$ as a function of temperature at 75 GPa is shown in the inset of Fig. 3.

Above calculation did not include phonon contribution. It is well known that the elastic constants can be affected by lattice vibrations, and thermal motion of ions is a common mechanism for material softening at high enough temperatures. Therefore, inclusion of phonon contribution might change the conclusion we just obtained above, especially at very high temperatures. In order to check this possibility and to consolidate our conclusion further, we also calculate the high temperature lattice vibrational contributions by using the self-consistent *ab initio* lattice dynamics (SCAILD) method,[53,54] which treats the anharmonicity in a self-consistent approach. From the self-consistent phonon dispersions, the lattice dynamical free energy can be computed from the phonon density of states, which was then used to correct the elastic moduli. Due to the huge computationally demanding of this method, and considering that the effect of lattice vibrations should be greater at high temperatures, we only calculate the phonon contribution with SCAILD method for Nb at 3000 K and 75 GPa. The obtained $C_{44}$ that includes both thermo-electron effects and phonon contribution at 3000 K and 75 GPa is shown in the inset of Fig. 3. Because the calculated lattice dynamical free energy always fluctuates when using the SCAILD method, an intrinsic feature of this algorithm, there is an uncertainty about ±10 GPa in the obtained $C_{44}$. From Fig. 3, it is evident that phonon softens the metal slightly. Its magnitude, however, is much smaller than the thermo-electronic contribution. Therefore, inclusion of phonon contribution cannot change the physical picture and our conclusion qualitatively. Because anharmonic phonon calculation is very computationally demanding, we will ignore its contribution in below discussion. This approximation should not affect our general conclusion qualitatively.

The shear anisotropy factor $A$ of Nb in BCC phase as a function of pressure at various electronic temperatures is plotted in Fig. 4. We find that the calculated $A$ always increases with temperature, indicating that the anisotropy of Nb gradually diminishes as the temperature increased. Its overall behavior is analogous to $C_{44}$, with the anomaly in the anisotropy occurs at 0~150 GPa and 275~400 GPa. Therefore the pressure-induced shear anisotropy in the BCC phase mainly originates from the elastic modulus $C_{44}$, rather than $C'$.

By concluding this subsection, we point out that the change of the shear anisotropy factor $A$ in the pressure range of 0~60 GPa with electronic temperature is not monotonic, as shown in Fig. 4. The value at zero Kelvin is slightly greater than that of 1000 K when the pressure is less than 40 GPa. This phenomenon is quite strange and will be discussed in details in the next subsection.

### E.  Charge transfer induced anomalies

As discussed above, Nb shows strong mechanical anomalies at high pressure. One might attempt to interpret this anomalous behavior using FS nesting, which appears in all of V, Nb, and Ta. However, the FS nesting ends at 75 GPa in Nb,[20] thus cannot account for the anomalies beyond that pressure. Also, it cannot explain the bizarre phenomenon of anisotropy intersection as shown in Fig. 4. In order to understand the electronic structure origination of these anomalies, we further study the phase stability of Nb at high pressure under various chemical environments. Charge-transfer at surface or interface could induce structural rearrangement,[55-57] and this electron accepting/donating will shift the Fermi level. Our previous study indeed found that the Fermi level has a remarkable impact on the phase stability of V.[18] By using the same method as employed in Ref. 18, we intendedly adjust the position of Fermi level to investigate the charge transfer effect on the shear anisotropy and the phase stability of Nb. In Fig. 5, we provide the enthalpy difference with respect to the BCC phase as a function of the rhombohedral deformation parameter $\delta$ for different charge transfer magnitude $\Delta$ at a selected pressure of 39 GPa. At this pressure, the enthalpy variation along the $RH_1$ distortion becomes the most flat when $\Delta = 0$ (that is, the enthalpy of $RH_1$ phase is mostly close to that of the BCC structure). Here $\Delta$ is defined as the total charge percentage being added to/removed from the system. As can be seen, the BCC phase is the only stable structure in the neutral system. When the Fermi level is pushed down by removing electrons, the RH phases gradually become stable, whereas the BCC phase is favored if the Fermi level is shifted up. In addition, our calculation indicates that the $RH_1$ phase is more stable than the BCC structure when $\Delta = -1.15\%$, and the largest stability of this phase is attained when $\Delta = -2.69\%$ at 39 GPa. However, further shifting down the Fermi level destabilizes the $RH_1$ phase. When $\Delta < -3.92\%$, the BCC becomes the favored phase again. A similar phase transition also occurs when the distortion parameter $\delta$ is positive, where the $RH_2$ phase is obtained. Details about this transition can be found in Supplementary Material.

This charge-transfer analysis reveals that there are two *underlying* RH phases in Nb, though they do not manifest explicitly in a neutral environment. The flat energy variation along the rhombohedral distortion path at $\Delta = 0$ indicates that the anomalous behavior of Nb at 0 K and 0~60 GPa as shown in Fig. 4 may originate in the combined contributions from the $RH_1$ and $RH_2$ deformations. In order to verify this assumption, we investigate the relationship between the $C_{44}$ deformation path and the $RH_1$ and $RH_2$ deformation paths. The corresponding elastic moduli as a function of pressure are plotted in Fig. 6. It can be seen that the elastic moduli of $RH_1$ and $RH_2$ deformations intersect at about 38 GPa. This is the same as that shown in Fig. 4. Moreover, it should be noted that the arithmetic average of them is consistent with the variation of $C_{44}$, suggesting that $C_{44}$ should contain some contributions from both $RH_1$ and $RH_2$ deformations. This can be easily understood from Fig. 7, which illustrates the relationship of these paths. It is evident that the $C_{44}$ deformation drives the BCC structure to transform into a monoclinic structure by the shear strain along the $[100]$ and $[\bar{1}00]$ directions. From that distorted structure, the $RH_1$ and $RH_2$ structure are obtained by shuffling just the $(001)$ plane along the $[\bar{1}\bar{1}0]$ or $[110]$ direction, respectively. From this we can conclude that the anomalous behavior of Nb at 0 K and 0~60 GPa mainly originates from the underlying $RH_1$ and $RH_2$ transition, the same softening mechanism as observed in vanadium. This unexpected connection is very interesting, and it is the slightly different response of $RH_1$ and $RH_2$ to thermo-electrons that leads to the intersection as manifested in Fig .6.

In order to acquire a more comprehensive understanding of the distortion mechanism, we further investigate the band structures of Nb in both $RH_1$ and BCC phase at 39 GPa for various $\Delta$, the results are plotted in Fig. 8. It can be seen that the band structure is just shifted up slightly as $\Delta$ decreased. When comparing the band structure of $RH_1$ to that of BCC, we find that the rhombohedral distortion splits the levels with $t_{2g}$ symmetry at $\boldsymbol{\Gamma}$ point. This effect is very similar to the Jahn-Teller distortion, and is consistent with the results of V as reported by Landa *et al.*[15] and Ohta *et al.*[58] It should be noted that all Fermi levels are well positioned in the pseudo-gap opened by the splitting of the $t_{2g}$ states at $\mathbf{\Gamma}$ point, even though the transition of BCC→RH actually does not take place when $\Delta = 0$. This intriguing phenomenon suggests that other factors besides the splitting of the $t_{2g}$ (such as *s-d* hybridization or electric quadrupole interaction) might also be involved in the charge-transfer induced BCC→RH phase transformation.

Furthermore, in order to understand the connection of the distortion mechanism between Nb and V. We also calculate the band structures of V and Nb in the BCC and $RH_1$ phases at 126 GPa and 39 GPa with $\Delta = 0$, respectively. The results are plotted in Fig. 9. At 126 GPa and 39 GPa, the $RH_1$ phase of V and Nb attains their greatest stability, respectively. It is evident that the band structures of Nb and V in both the BCC and $RH_1$ phases are very similar to each other, especially at the vicinity of the Fermi level. This remarkable similarity unequivocally denotes that the softening mechanism at the respective pressure should be the same for both Nb and V. The only difference is that the BCC→RH transition really occurs in V, but the driven force is much smaller in Nb so that hinders the occurrence of the phase transition. This analysis also demonstrates that the transition could explicitly occur in Nb if one can temporally remove the electrons from the valence band by using, say, high energy photon-excitations, or putting it to a surface or interface with appropriate chemical treatments where large charge transfer can occur.

## IV. DISCUSSION

An intriguing question is that whether the above mechanism can interpret the newly discovered anomalous softening of $C_{44}$ at 275~400 GPa in Nb? The answer is negative, because we find that the charge-transfer has almost no impact on the RH stability at this pressure range. That is, the underlying $RH_1$ and $RH_2$ deformation is not relevant in this pressure range. Other factors should take the responsibility for the $C_{44}$ softening at 275~400 GPa. In order to elucidate that mechanism, we calculate the band structure and the Fermi surface of Nb around this pressure. The results are plotted in Fig. 10, which reveals that Nb undergoes an ETT on the Fermi surface at a pressure of approximately 300 GPa. Since there are no FS nesting and Jahn-Teller splitting in this pressure range, this pressure-induced ETT should be the cause for the anomalous behavior of Nb at 275~400 GPa.

It is necessary to point out that the variation of the elastic modulus $C_{44}$ as a function of pressure for V, Nb, and Ta are not the same, even they have almost the same valence electronic configuration. In order to obtain a comprehensive understanding of the connection between them, we carefully calculate and compare the band structures of them in the BCC phase from 0 to 400 GPa (the figures can be found in Supplementary Material). We find that for all of V, Nb, and Ta, there are actually two ETTs, both along the $\mathbf{\Gamma \rightarrow H}$ direction. The first ETT occurs at about 300, 110, and 280 GPa in V, Nb, and Ta, respectively. The second ETT occurs at 300 GPa in Nb and at about 600 GPa in both V and Ta. On the other hand, the band Jahn-Teller effect becomes effective at 250, 100, and 240 GPa in V, Nb, and Ta, respectively. The FS nesting in these three metals covers a range of 0~275, 0~75, and 0~247 GPa, respectively. It is thus obvious that the softening mechanism for V at 20~300 GPa, for Nb at 20~150 GPa, and for Ta at 140~250 GPa should be the same. They all are induced by a combination of the contributions from ETT, Jahn-Teller effect, and FS nesting, which together leads to a huge softening. This, however, does not necessarily result in the RH distortions, and in Ta we find no sign of this deformation by charge transfer analysis. Furthermore, Ta does not follow the thumb rule that for elements in the same group the phenomenon in the light element at high pressures should occur at lower pressures in the heavier elements. We also notice that in both V and Nb $C_{44} < C'$, but it becomes $C_{44} > C'$ in Ta. This could be due to the semi-core $4f$ electrons that are absent in both V and Nb. In this sense, the RH distortion might also relate to the electric quadrupole interaction, which is stronger in V and Nb. In addition, it is quite unusual that a second anomalous softening of $C_{44}$ is predicted in Nb at 275~400 GPa, which are absent from the same group members V and Ta at the same pressure range. As mentioned above, the second pressure-induced ETT should be responsible for this anomalous behavior. Since the Fermi level gradually shifts up as the pressure increases, V and Ta also undergo this ETT at higher pressures. We thus predict that V and Ta should have the same softening at high pressure of 600 GPa (see Fig. S6 in the Supplementary Material). This is a quite interesting prediction, but beyond the scope of this work and we will investigate them in the future study.

## V. CONCLUSIONS

In summary, a thorough and comprehensive theoretical study of Nb at high pressure and high temperature has been carried out with first-principles calculations based on density functional theory. Our study demonstrated that the contribution from the $RH_1$ and $RH_2$ deformations leads to the $C_{44}$ and $C'$ softening in Nb at 20~150 GPa. Furthermore, we found another previously unknown anomalous softening of $C_{44}$ at 275~400 GPa, and a second pressure-induced ETT is responsible for it. Based on this, a unified physical interpretation for all high pressure anomalies of group VB is obtain. With this picture, we further predicted that a new softening will occur in both V and Ta at a pressure about ~600 GPa. Moreover, we studied the thermo-electron effects on the elastic properties at finite temperatures. The results revealed a peculiar but strong increase in the shear modulus $C_{44}$ with increasing electronic temperature.

We also showed that the shear modulus and Young's modulus of Nb are highly anisotropic, while the linear compressibility is isotropic. At 60 GPa, the anisotropy in the shear modulus reaches the maximum. But at finite temperatures, this anisotropy gradually diminishes with the temperature increases. Our calculations clearly revealed the electronic structure originations of the anomalies in Nb, and its close connection to V and Ta. Nonetheless, the weak softening of Ta at ~80 GPa as reported by Landa *et al*.[59] and Jing *et al*.[60] cannot be accounted for by these mechanisms (there is no ETT or RH distortion), which is still an open question and require further investigation.

## SUPPLEMENTARY MATERIAL

See supplementary material for (a) Fermi level shift and charge transfer analysis for $RH_2$ phase, and (b) elastic anisotropic characters of Nb.

## ACKNOWLEDGMENTS

This work is supported by the NSAF under Grant Nos. U1730248 and U1430117,

the National Natural Science Foundation of China under Grant Nos. 11672274 and 11602251, the Fund of National Key Laboratory of Shock Wave and Denotation Physics of China under Grant No. 6142A03010101, the CAEP Research Project under Grant No. 2015B0101005, and the Science Challenge Project under Grant No. Tz2016001.

**Table I.** The calculated primitive cell volume $V_0$ (Å$^3$), bulk modulus $B_0$ (GPa) and its pressure derivative $B_0'$ of Nb at zero Kelvin and zero pressure, together with the available experimental and theoretical data.

|    | Reference | $V_0$ | $B_0$ | $B_0'$ |
|----|-----------|-------|-------|--------|
| Nb | This study | 18.11 | 169.79 | 3.61 |
|    | Theor.[16] | 18.32 | 174 | 3.85 |
|    | Expt.[31] | 17.98 | — | — |
|    | Expt.[9] | 17.98 | 168.4 | 3.43 |

**Table II.** The calculated elastic constants $C_{11}$, $C_{12}$, and $C_{44}$ (in GPa) of Nb in BCC phase at 0 GPa and 0 K, together with the available experimental and theoretical data.

|    | Reference | $C_{11}$ | $C_{12}$ | $C_{44}$ |
|----|-----------|----------|----------|----------|
| Nb | This study | 246.99 | 131.19 | 19.77 |
|    | Theor.[15] | — | — | 25.3 |
|    | Theor.[16] | 247.0 | 138.0 | 10.3 |
|    | Theor.[19] | 247.8 | 133.1 | 17.2 |
|    | Expt.[33] | 246.5 | 134.5 | 28.7 |

**Table III.** Calculated anisotropic indices of Nb in BCC phase under various pressures at zero Kelvin.

| $P$ (GPa) | $A$ | $A_G$ | $A_U$ |
|-----------|-----|-------|-------|
| 0 | 0.34 | 0.13 | 1.31 |
| 10 | 0.30 | 0.16 | 1.39 |
| 20 | 0.29 | 0.17 | 1.42 |
| 30 | 0.27 | 0.19 | 1.47 |
| 40 | 0.25 | 0.21 | 1.53 |
| 50 | 0.21 | 0.26 | 1.70 |
| 60 | 0.16 | 0.34 | 2.02 |
| 70 | 0.18 | 0.31 | 1.89 |
| 80 | 0.25 | 0.21 | 1.54 |
| 90 | 0.31 | 0.15 | 1.36 |
| 100 | 0.38 | 0.11 | 1.25 |

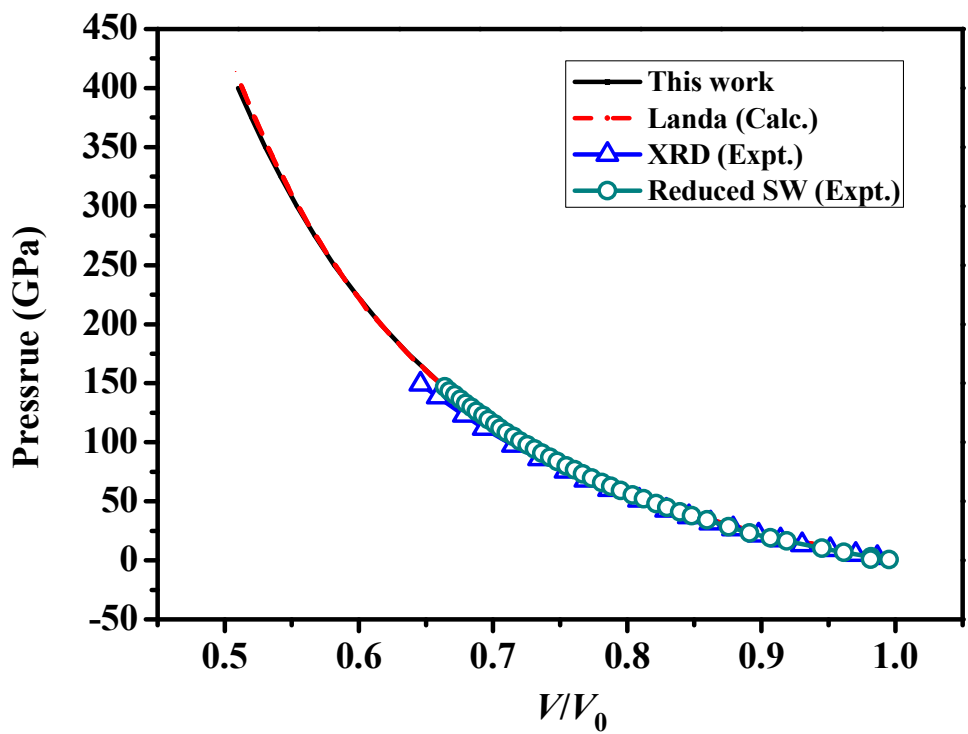

**Figure 1.** (Color online) Isothermal EOS (*P-V* curve) of Nb at zero Kelvin. The dashed line is the calculated result from Landa *et al.*[15] The static x-ray diffraction (XRD) data and shock wave (SW) data that have been reduced to *T* = 0 K are taken from Refs. 9 and 32, respectively.

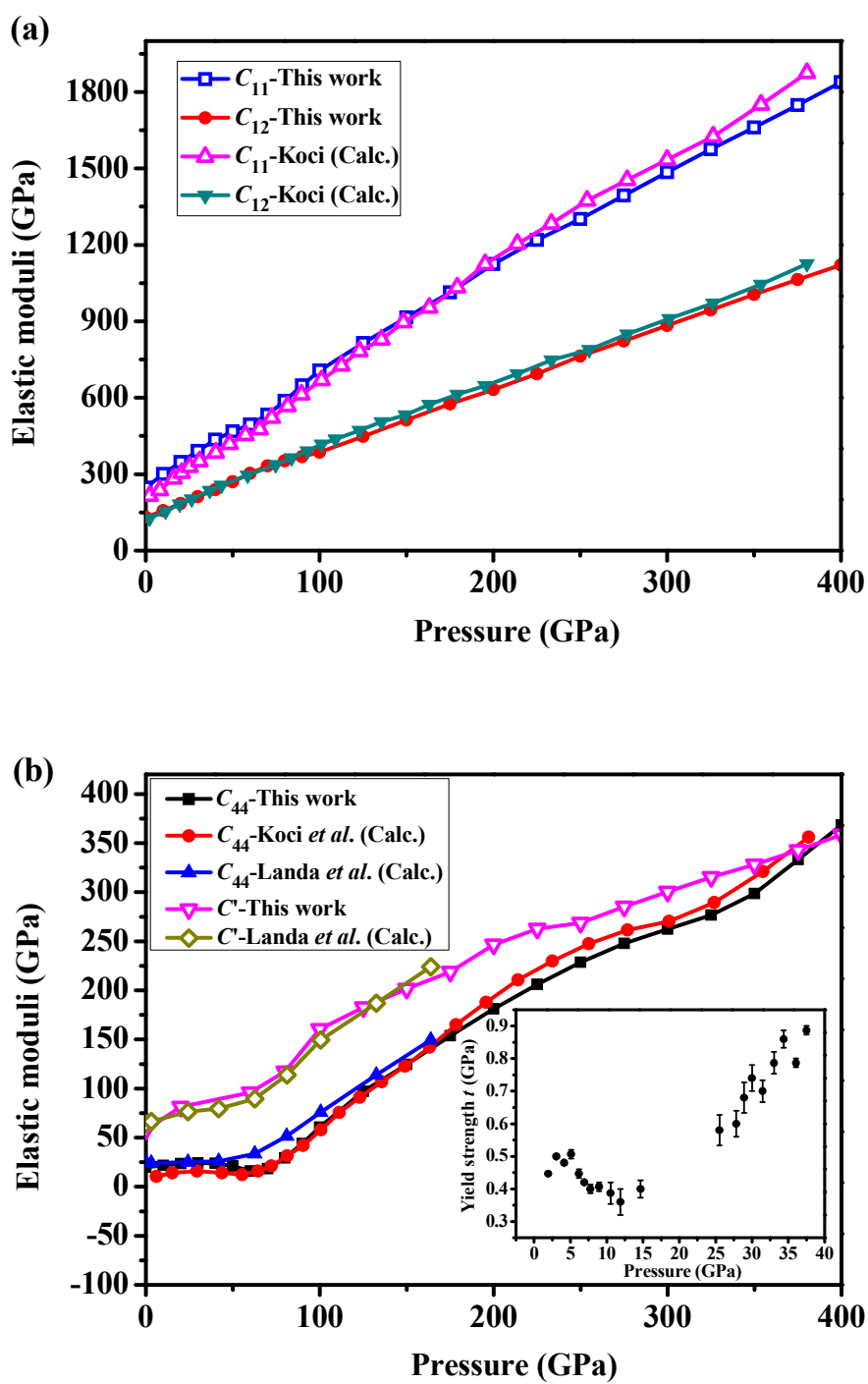

**Figure 2.** (Color online) The calculated elastic moduli (a) $C_{11}$ and $C_{12}$, (b) $C_{44}$ and $C'$ of BCC Nb as a function of pressure at zero Kelvin by comparison with theoretical data of Landa *et al.*[15] and Koči *et al.*[16] The inset of (b) shows the yield strength of Nb as function of pressure measured by Singh *et al.*[12]

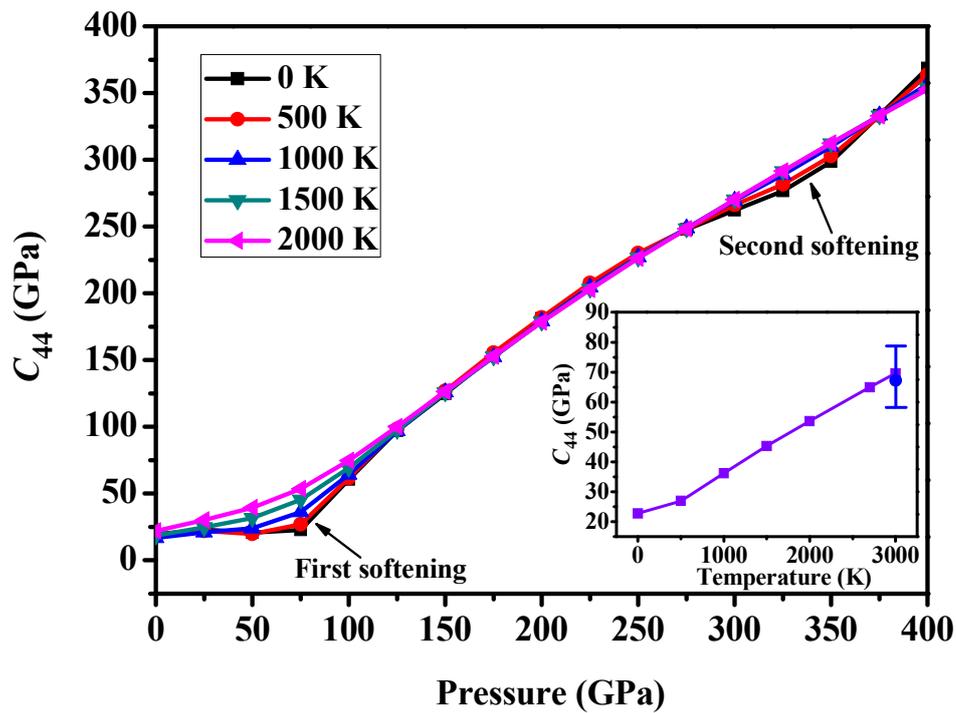

**Figure 3.** (Color online) Calculated shear elastic constant $C_{44}$ of Nb in BCC phase as a function of pressure at different electronic temperatures. Inset: Variation of $C_{44}$ as a function of electronic temperature at a pressure of 75 GPa. The blue solid circle with error bar denotes the $C_{44}$ that includes both thermo-electron effect and phonon contribution at 3000 K and 75 GPa.

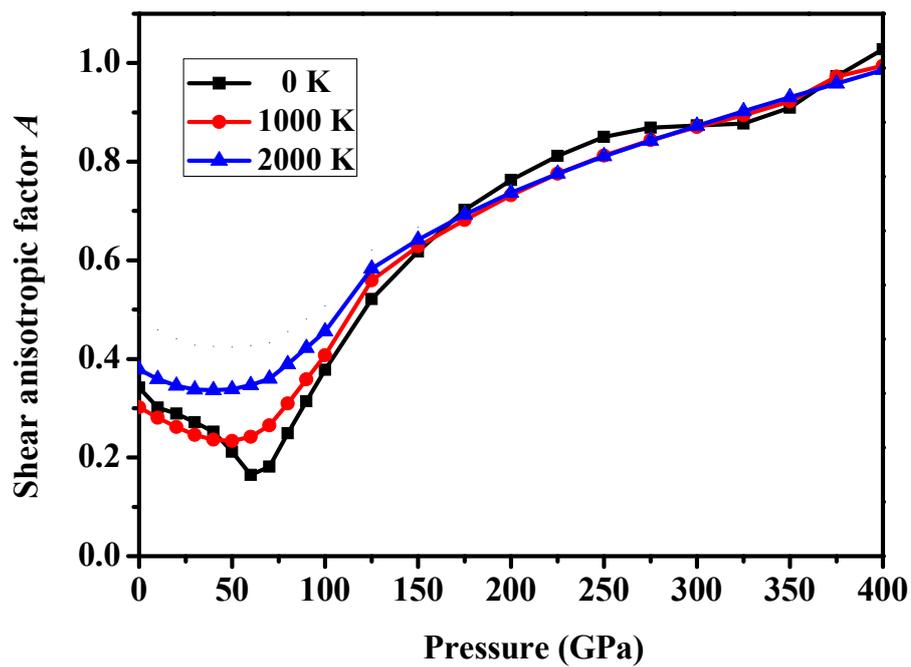

**Figure 4.** (Color online) Variation of the shear anisotropy factor *A* of Nb in BCC phase

as a function of pressure at different electronic temperatures.

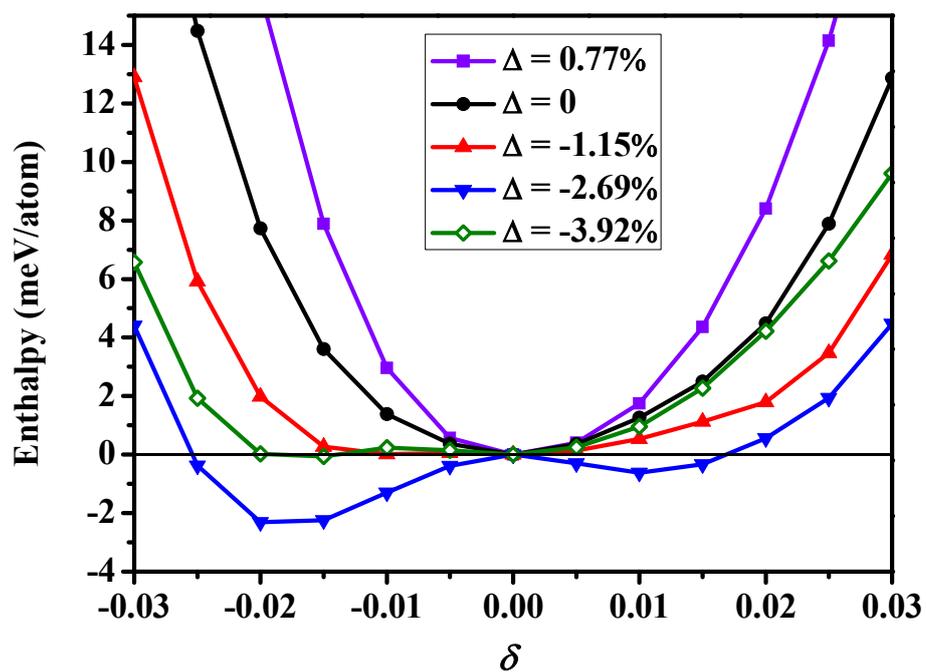

**Figure 5.** (Color online) Variation of the enthalpy difference with respect to the BCC phase as a function of the rhombohedral deformation parameter $\delta$ for different charge transfer $\Delta$ at a pressure of 39 GPa.

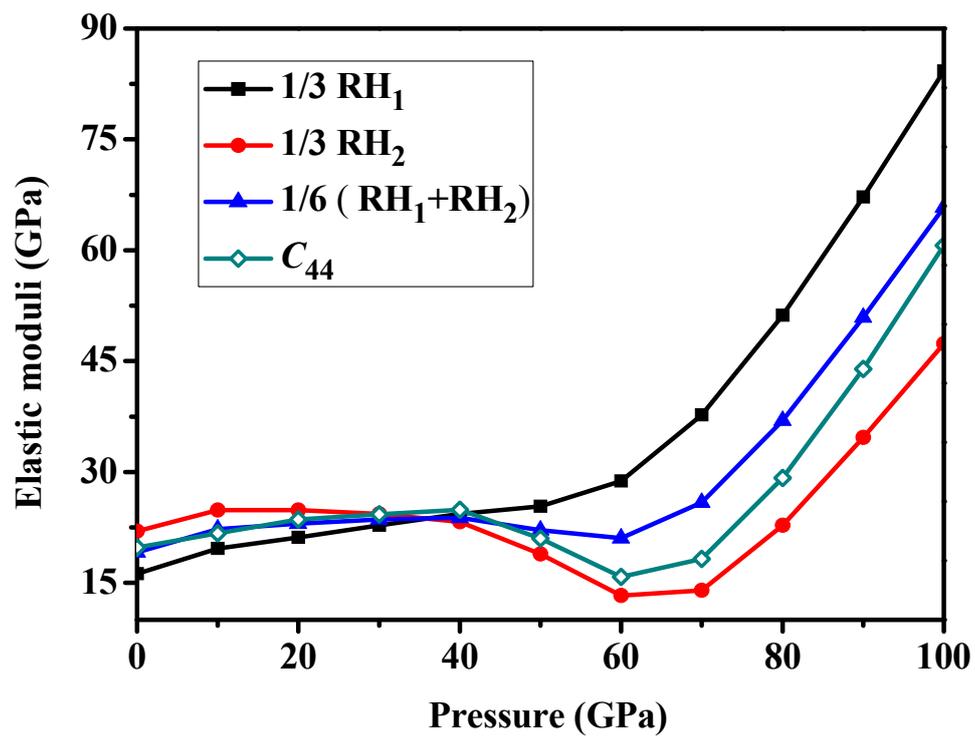

**Figure 6.** (Color online) The elastic moduli corresponding to the $RH_1$, $RH_2$, and $C_{44}$ deformations as a function of pressure at zero Kelvin.

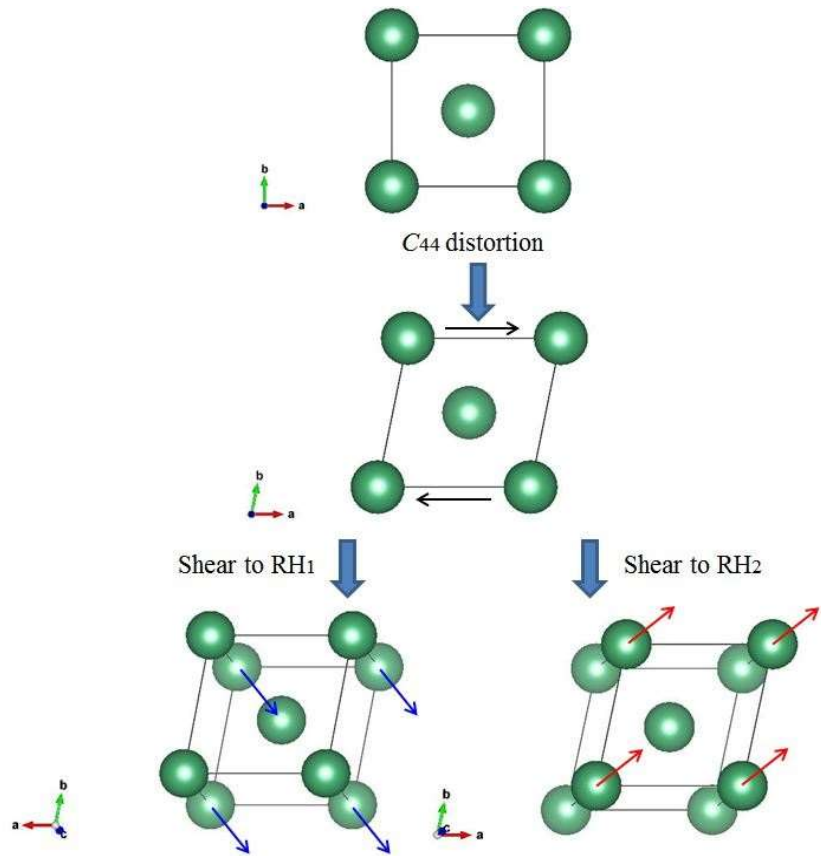

**Figure 7.** (Color online) The transformation path of Nb from the BCC structure to the

RH structures.

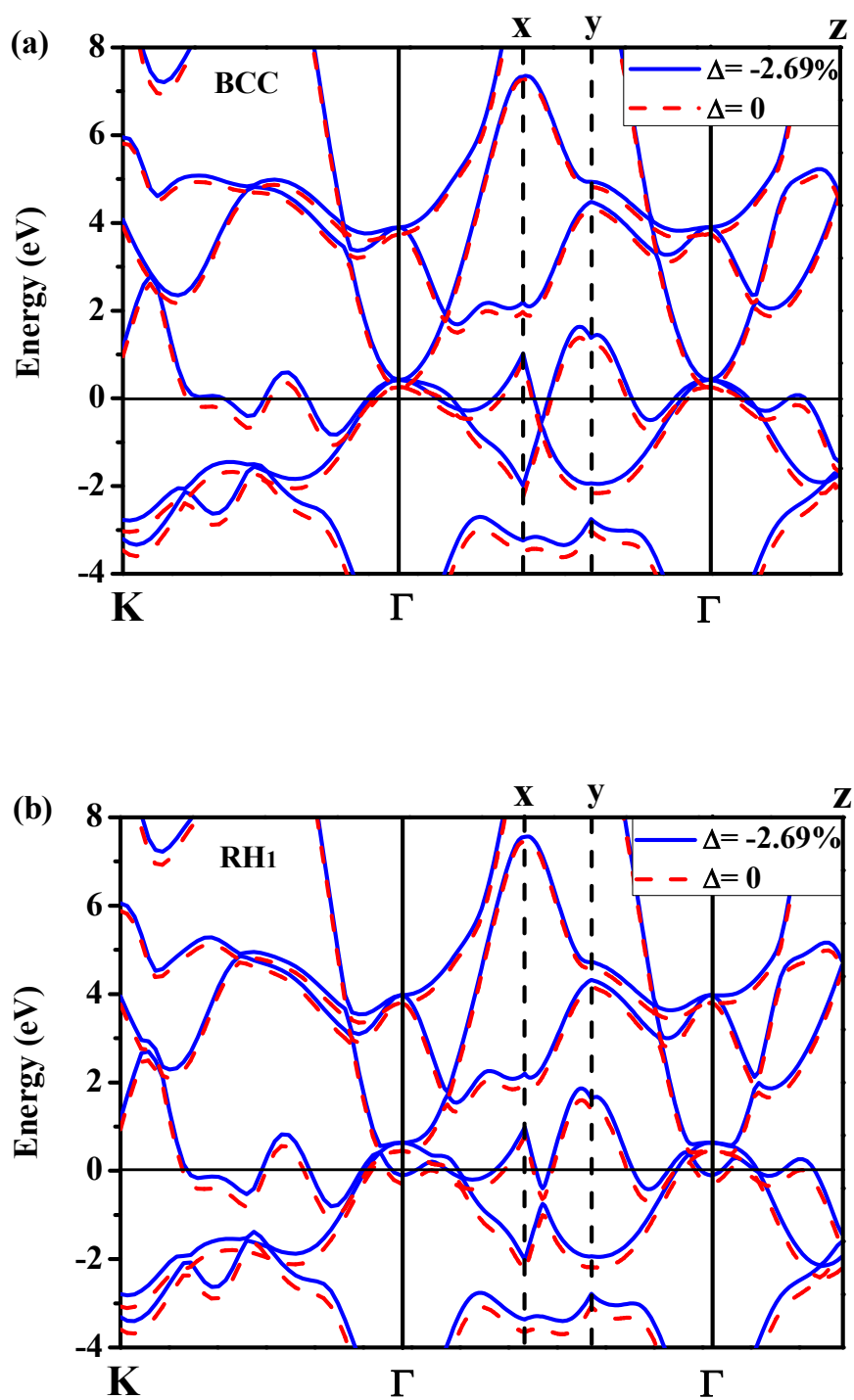

**Figure 8.** (Color online) Band structures of Nb at 39 GPa for different $\Delta$: (a) BCC phase,

(b) RH$_1$ phase. Here, the **x**, **y**, and **z** denote the high symmetry points of (1/4, -1/4, 1/2),

(1/8, 1/8, 3/8), and (3/8, -1/8, 3/8), respectively.

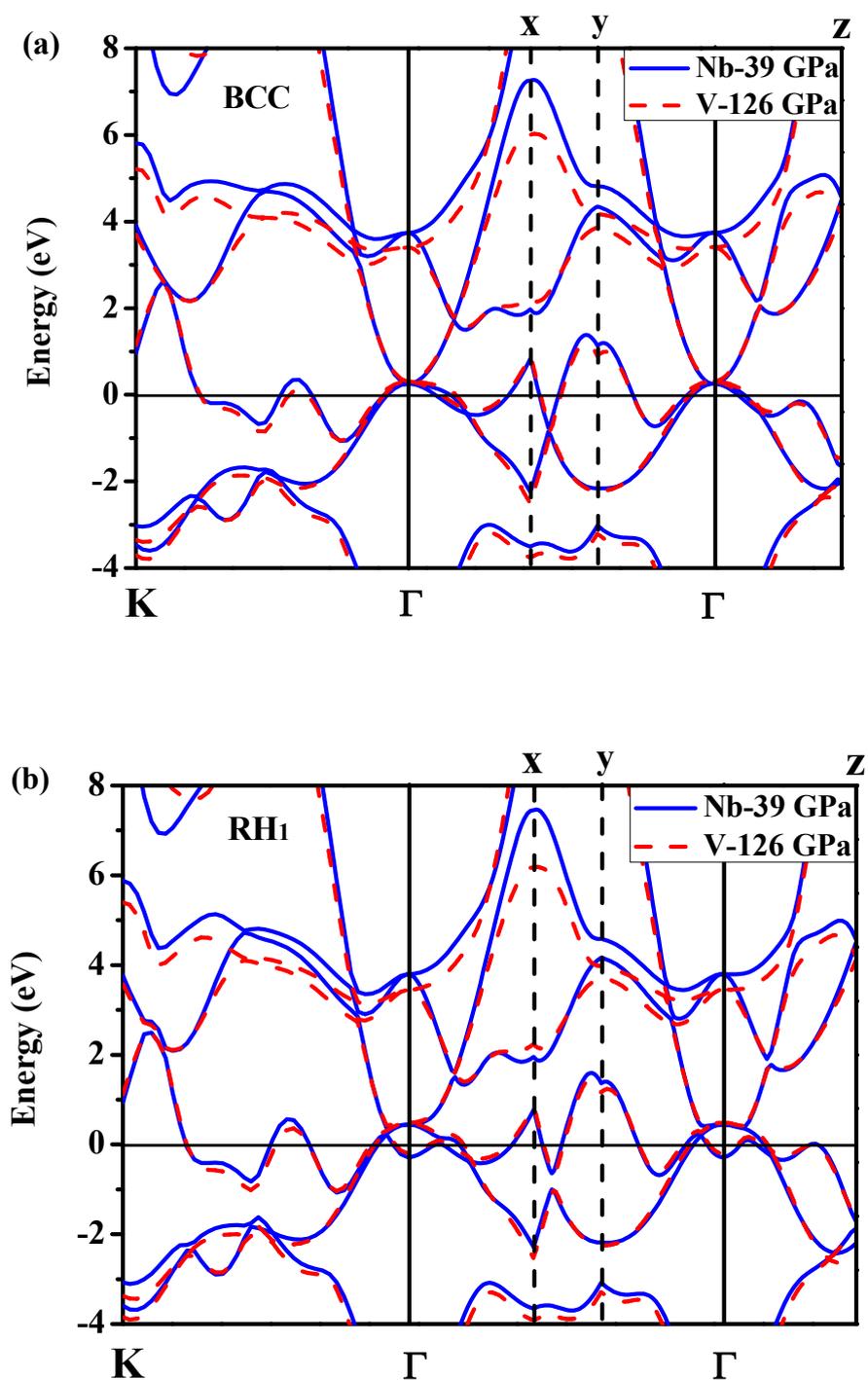

**Figure 9.** (Color online) Band structures of V (at 126 GPa) and Nb (at 39 GPa) when $\Delta = 0$: (a) BCC phase, (b) $RH_1$ phase. Here, the **x**, **y**, and **z** denote the high symmetry points of (1/4, -1/4, 1/2), (1/8, 1/8, 3/8), and (3/8, -1/8, 3/8), respectively.

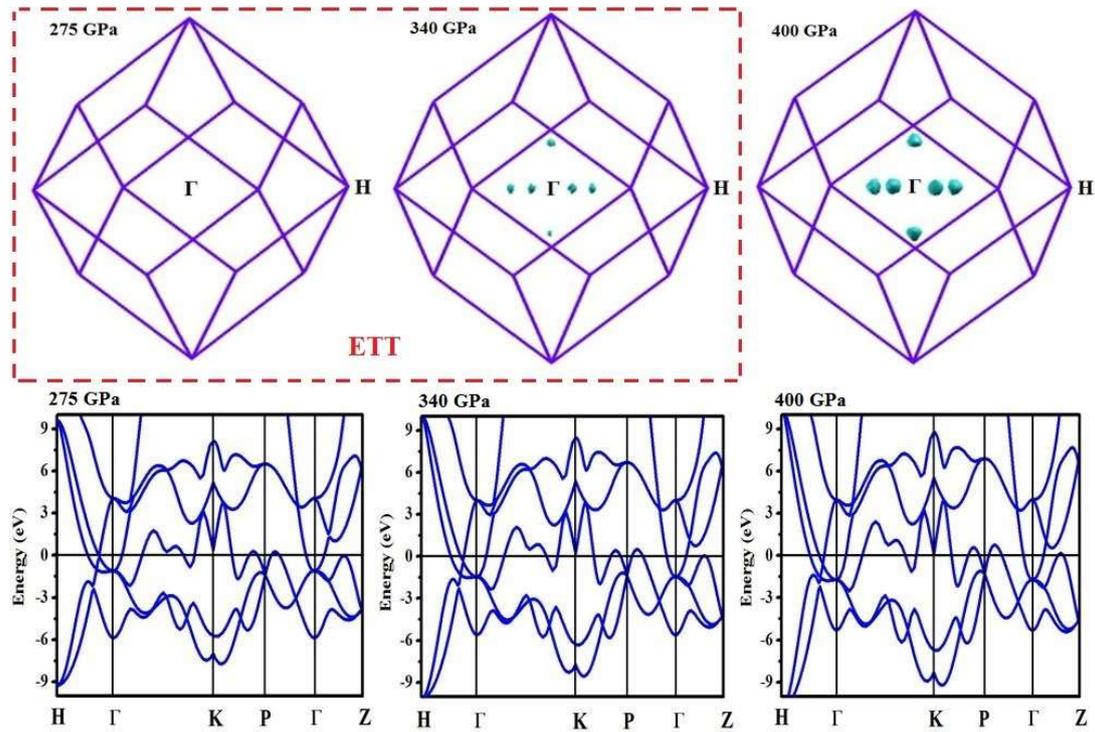

**Figure 10.** (Color online) The calculated Fermi surface (above) and band structure (below) of BCC Nb at different pressures. The ETT occurs when a new band (the band 9) appears along the **Γ→H** directions between 275 and 340 GPa, which changes the topology of the Fermi surface.

Supplementary information for

# First-principles investigation of elastic anomalies in niobium at high pressure and temperature

Yi X. Wang,[1,2] Hua Y. Geng,[2,*] Q. Wu,[2] Xiang R. Chen,[1,*] and Y. Sun[2]

[1] *Institute of Atomic and Molecular Physics, College of Physical Science and Technology, Sichuan University, Chengdu 610064, China;*

[2] *National Key Laboratory of Shock Wave and Detonation Physics, Institute of Fluid Physics, CAEP, Mianyang 621900, China.*

*e-mail: s102genghy@caep.ac.cn; xrchen@scu.edu.cn*

## A. Fermi level shift and charge transfer analysis for $RH_2$ phase

In Fig. S1, we show the enthalpy difference with respect to the BCC phase as a function of the rhombohedral deformation parameter $\delta$ for different $\Delta$ at a selected pressure of 61 GPa. At this pressure, the enthalpy variation along the $RH_2$ distortion is the most flat when $\Delta = 0$. The same as $RH_1$ distortion, our calculation indicates that the $RH_2$ phase becomes more stable than the BCC structure when $\Delta = -0.62\%$, and the maximal stability of $RH_2$ phase is attained at $\Delta = -1.46\%$. Further shifting down the Fermi level will destabilize the $RH_2$ phase. When $\Delta < -2.35\%$, the $RH_2$ transforms back to the BCC phase again.

In order to deepen our understanding about the distortion mechanism, the variation of the band structures of Nb in $RH_2$ and BCC phases at 61 GPa for different $\Delta$ are calculated and showed in Fig. S2. It can be seen that the energy bands will slightly shift up with $\Delta$ decreasing. When comparing the band structure of $RH_2$ to BCC phase, we find that the rhombohedral distortion splits the levels with $t_{2g}$ symmetry at $\boldsymbol{\Gamma}$ and $\mathbf{H}$ points. Therefore, analogous to $RH_1$ deformation, the energy band splitting induced by this distortion is the same as the band Jahn-Teller effect, which also promote the BCC→$RH_2$ phase transition.

As mentioned in the main text, the BCC→RH transition in Nb and V should have the same mechanism. In order to verify this comprehension, we compare the band structures of V (at 211 GPa) and Nb (at 61 Pa) in both the BCC and $RH_2$ phase at $\Delta = 0$. The results are plotted in Fig. S3. At the selected pressure of 211 GPa, the $RH_2$ phase of V attains the greatest stability. It is evident that the band structures of Nb and V are very similar with each other at the vicinity of the Fermi level when they are in the same phase (BCC or $RH_2$), which confirms that the electronic structure origination of this phase transformation is the same for both V and Nb.

In Fig. S4, we plot the calculated band structures of V, Nb, and Ta in the BCC phase at different pressures. From them we can find that the Fermi level of V, Nb, and Ta are all gradually shifted up as the pressure increases. With the increasing pressure, the band Jahn-Teller effect becomes effective at about 250, 100, and 240 GPa for V, Nb, and Ta, respectively. Furthermore, the calculated Fermi surfaces in the BCC phase of V, Nb, and Ta at different pressures are shown in Fig. S5. It can be seen that the first ETT occurs in the **Γ→H** directions at 240~280, 75~100, and 225~250 GPa for V, Nb, and Ta, respectively. At 275~340 GPa, a second ETT occurs in the **Γ→H** directions for Nb, which results in the anomalous softening of elastic moduli $C_{44}$. If the pressure further increases, the same ETT will also occur in V and Ta at about 600 GPa (see Fig. S6). That is, the V and Ta also have a second anomalous softening in $C_{44}$ at higher pressures.

## B. Elastic anisotropic characters of Nb

In order to understand the elastic anisotropy more intuitively, we present the anisotropic characters of shear modulus at 60 GPa in Fig. S8, in which (a) and (b) represent the maximal and minimal shear modulus, respectively. It is evident that the shear moduli exhibit a high degree of anisotropy in different crystallographic orientations. In addition, we calculate two 3D surfaces by using the Eqs. (15) and (16) (see main text), in which the distance from the origin of the coordinates to the surface equals to the Young's modulus $E$ or the linear compressibility $\beta$ in a given direction. The results for Nb at 60 GPa are shown in Fig. S9. It is evident that though the shear deformation in Nb is highly anisotropic, the linear compressibility is strictly isotropic.

Nonetheless, the Young's modulus displays a strong anisotropy, with the largest value presents along the axis directions.

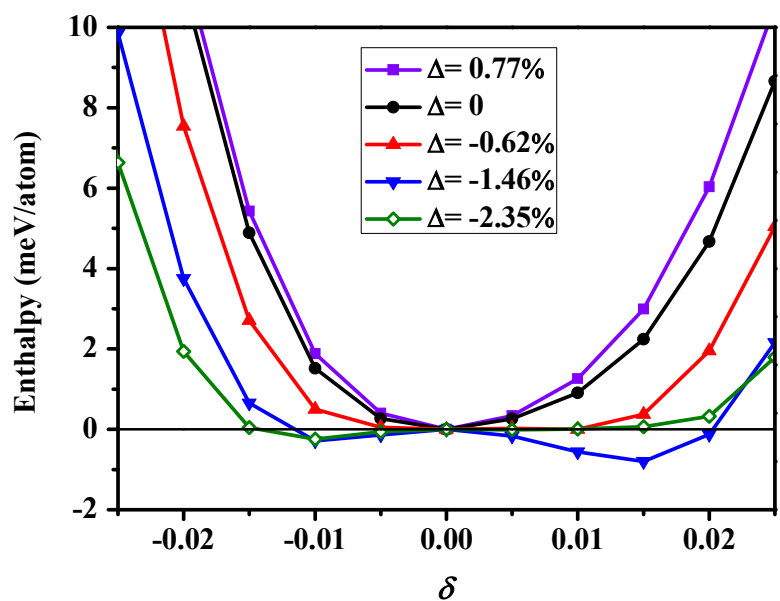

**Figure S1.** (Color online) Variation of the enthalpy difference with respect to BCC phase as a function of the rhombohedral deformation parameter $\delta$ for different $\Delta$ at 61 GPa.

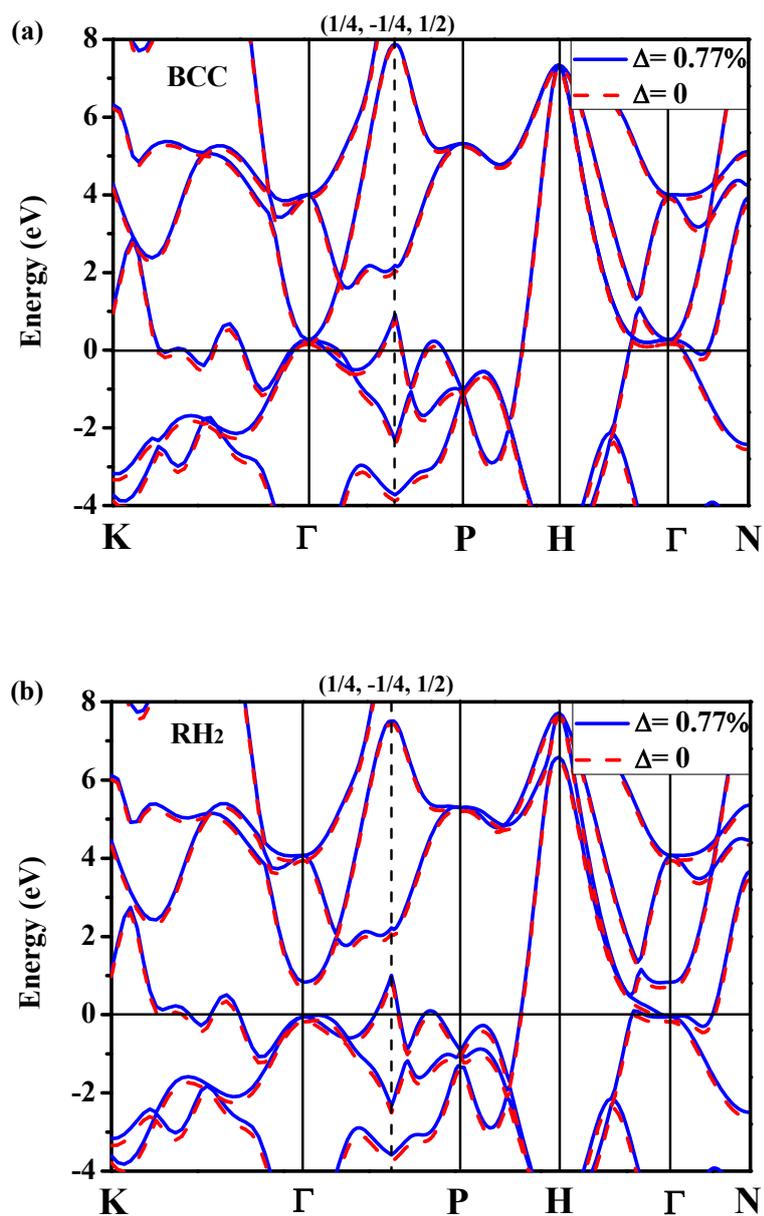

**Figure S2.** (Color online) Band structures of Nb at 61 GPa for different $\Delta$: (a) BCC phase, and (b) RH$_2$ phase.

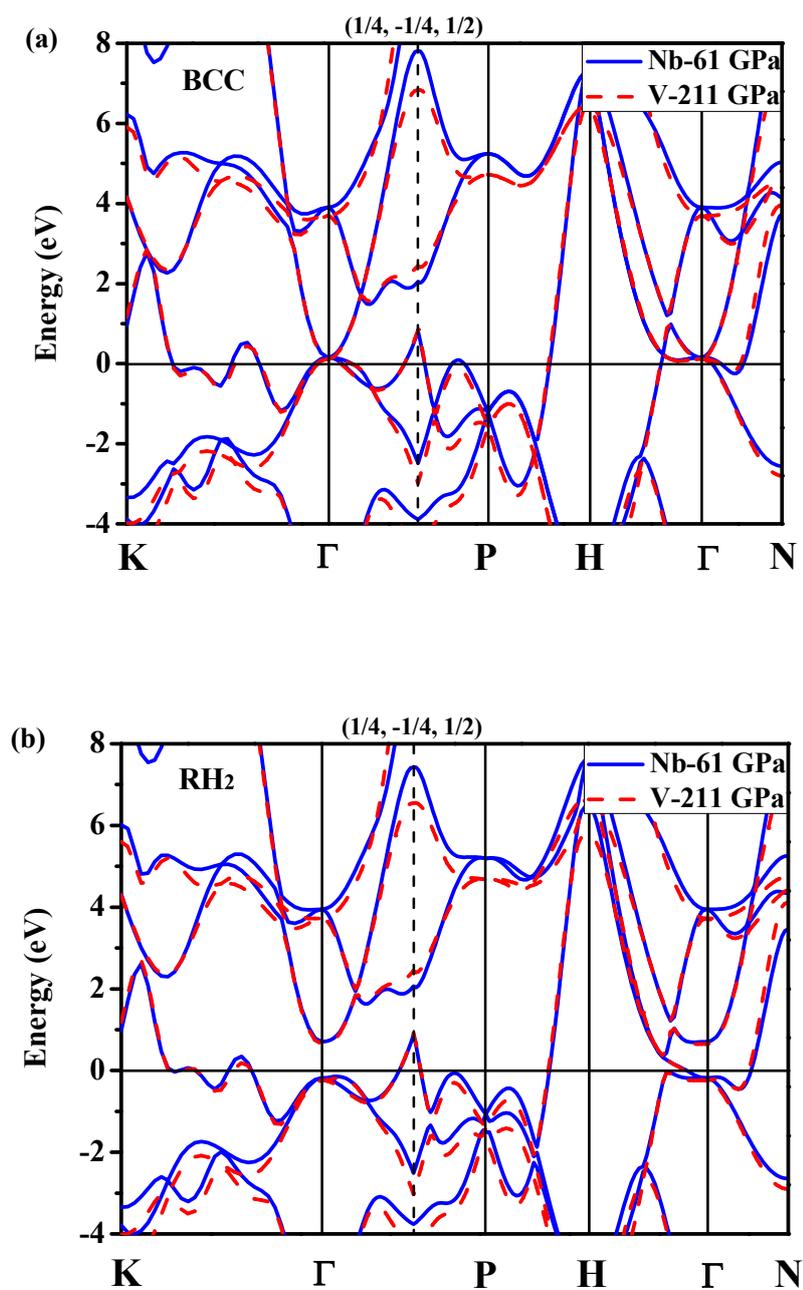

**Figure S3.** (Color online) Band structures of V (at 211 GPa) and Nb (at 61GPa ) when

$\Delta = 0$: (a) BCC phase, and (b) RH$_2$ phase.

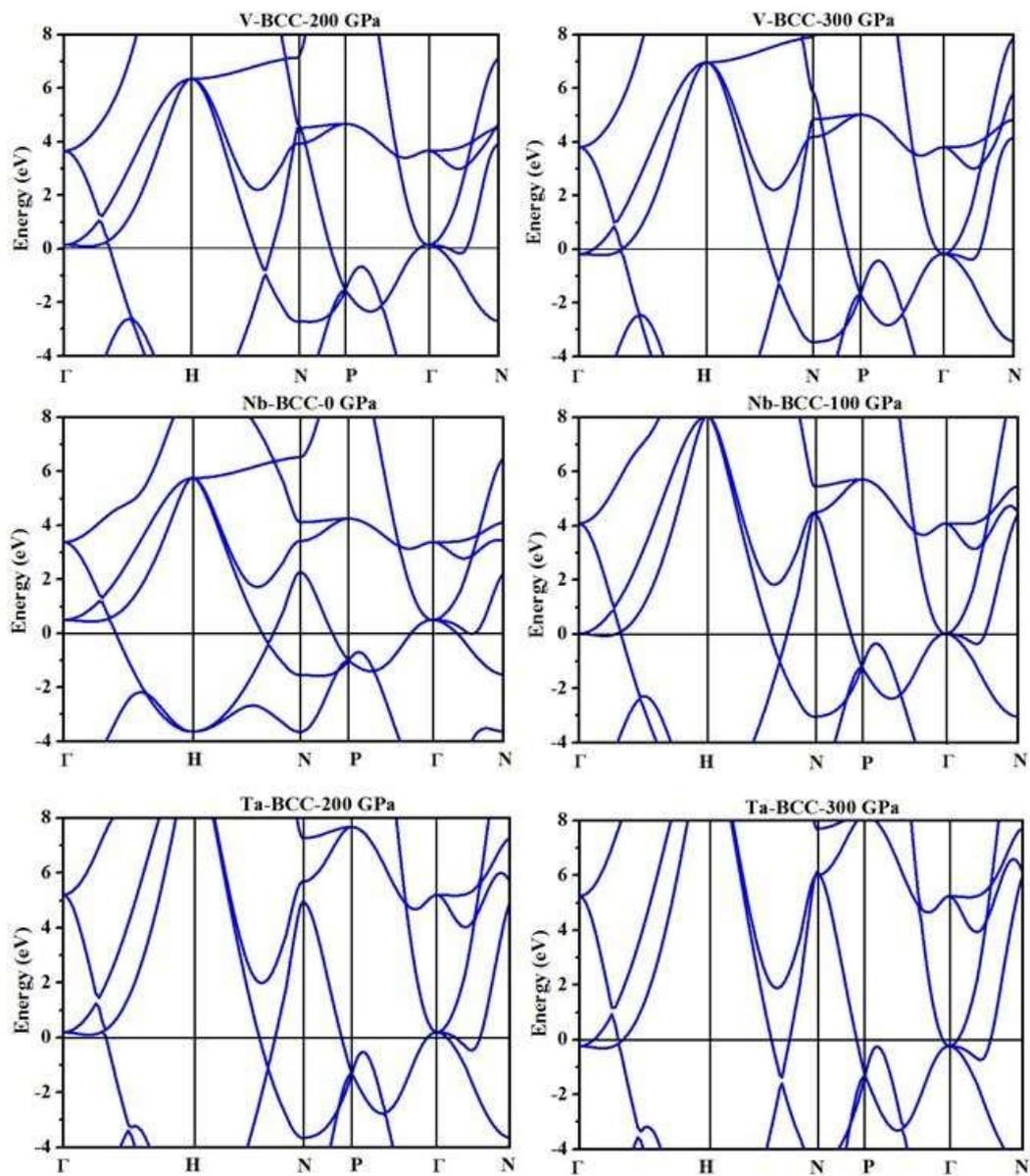

**Figure S4.** (Color online) Band structures of V, Nb, and Ta in BCC phase at various pressures.

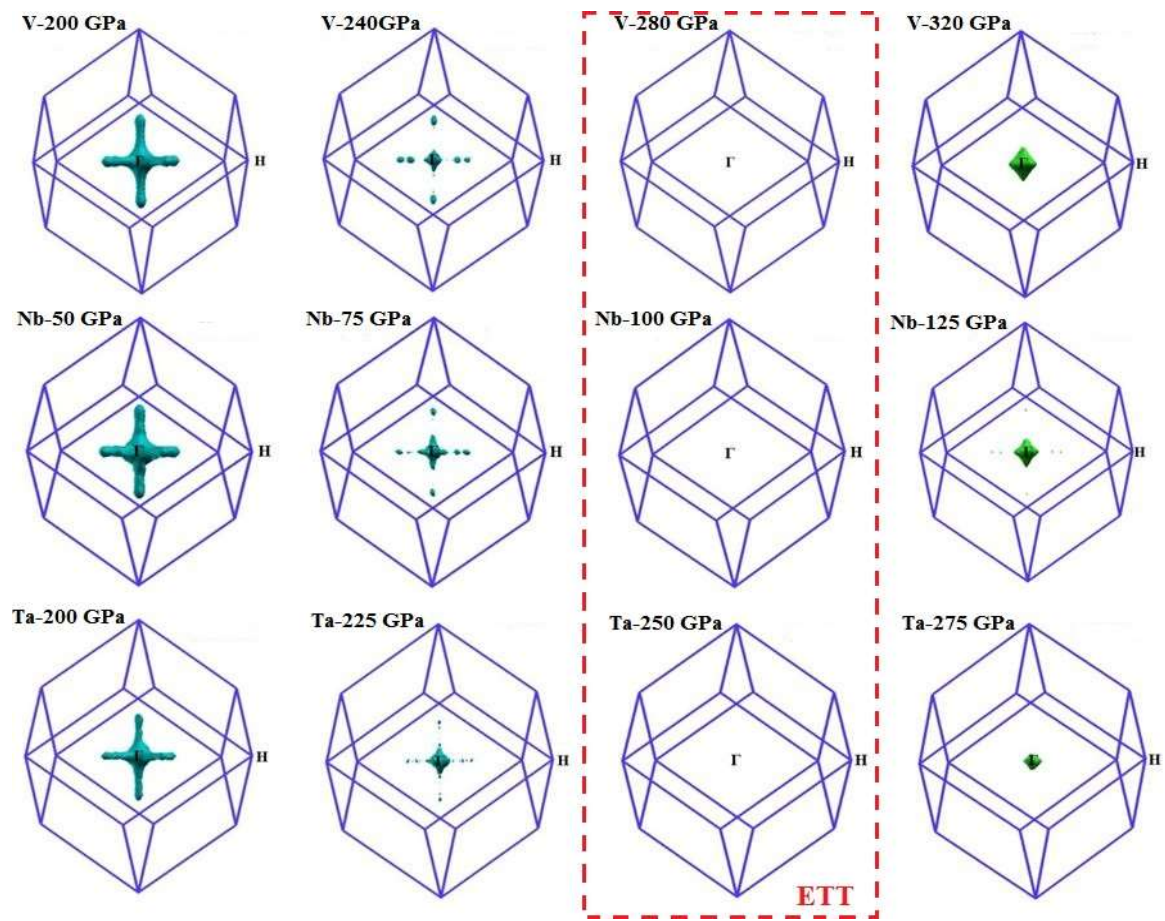

**Figure S5.** (Color online) Calculated Fermi surface of V, Nb, and Ta in BCC phase at various pressures. The change of the topology of the Fermi surface, and thus the ETT, is evident.

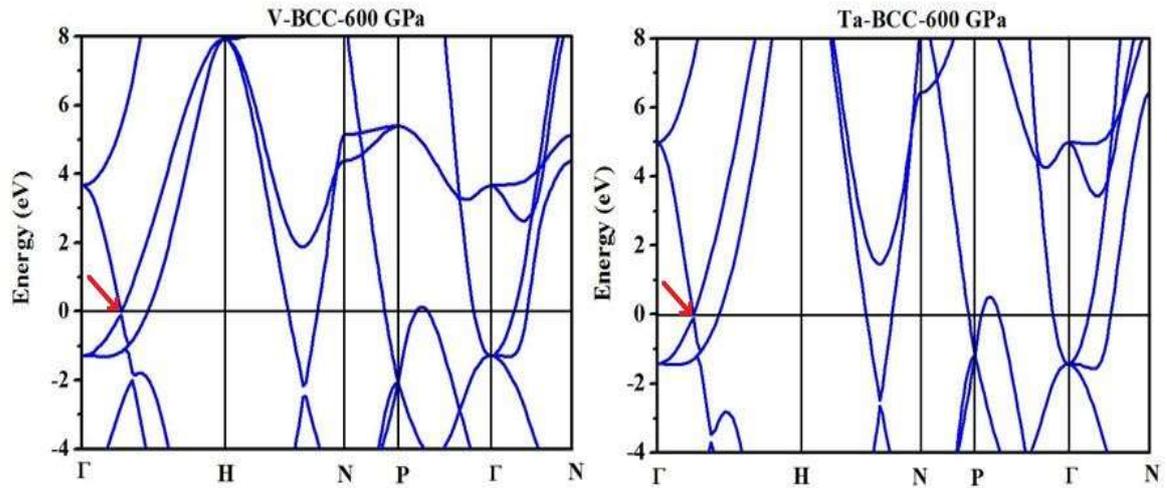

**Figure S6.** (Color online) The calculated band structures of V and Ta in BCC phase at 600 GPa, where a second ETT is predicted, as the arrow points out. See also Fig. 12 of the main text.

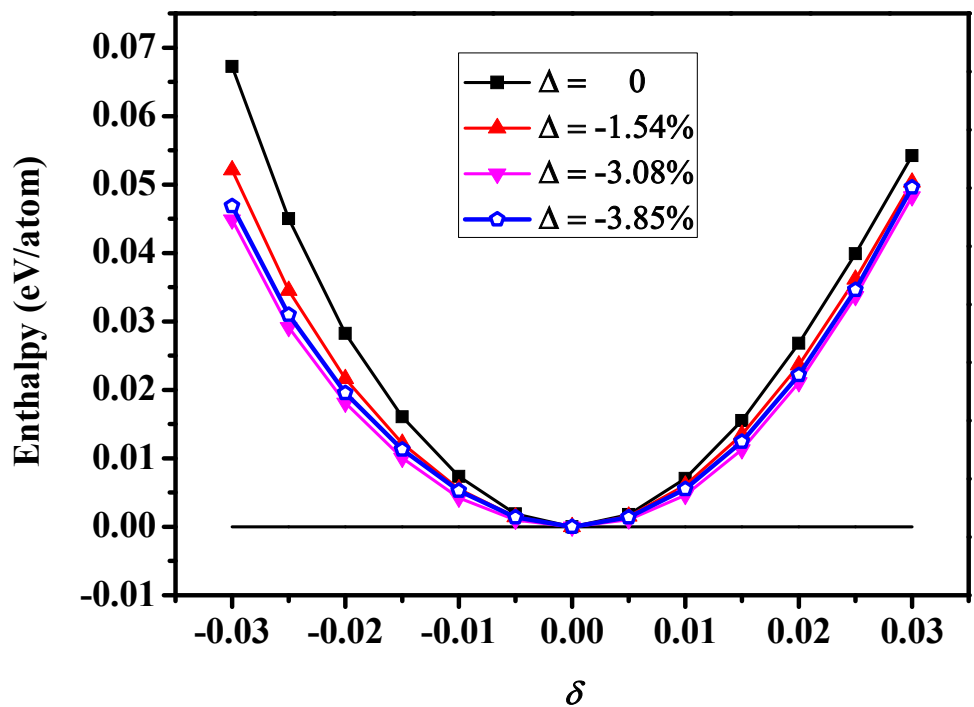

**Figure S7.** (Color online) Variation of the calculated enthalpy differences of Ta with respect to the BCC phase as a function of the rhombohedral deformation parameter $\delta$ for different charge transfer $\Delta$ at 80 GPa. A slightly softening due to change-transfer is observed. However, it is unclear whether this has a direct connection to the reported strength softening of Ta at ~80 GPa or not.

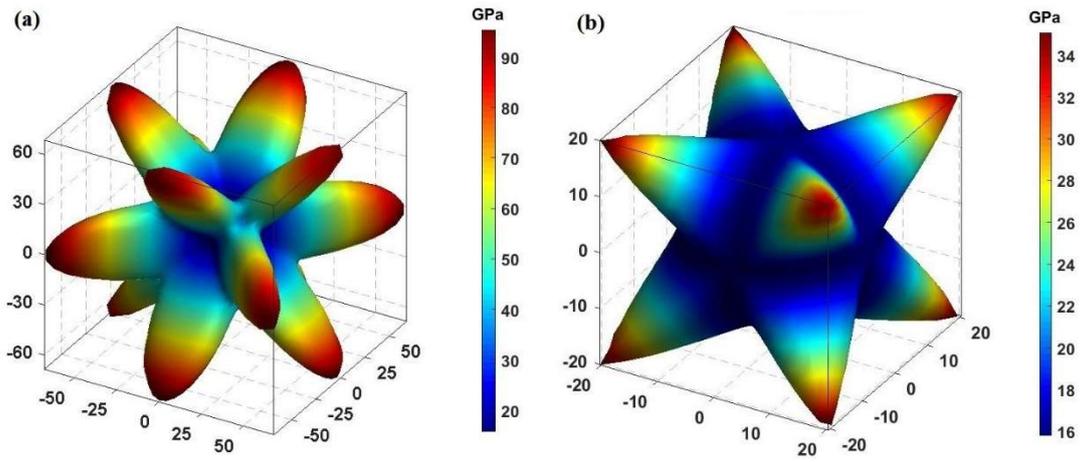

**Figure S8.** (Color online) The shear modulus anisotropic characters of Nb in BCC phase at 60 GPa, where (a) and (b) represent the maximal and minimal shear, respectively.

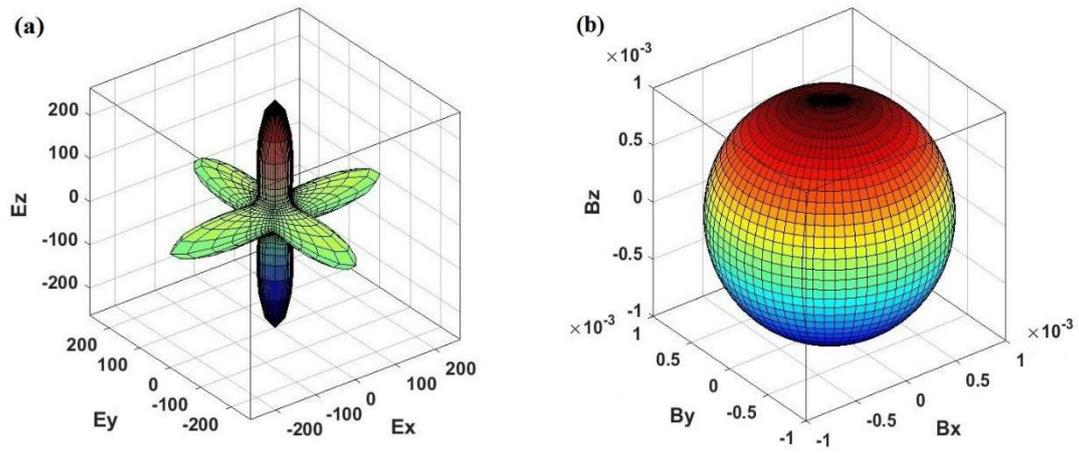

**Figure S9.** (Color online) The directional dependence of (a) Young's modulus (in GPa), and (b) linear compressibility (in GPa$^{-1}$) of Nb in BCC phase at 60 GPa.